\documentclass[useAMS,usenatbib]{mn2e}
\usepackage{amsmath}
\usepackage{times}
\usepackage{graphicx}
\usepackage{aas_symbols}
\usepackage{graphicx}
\usepackage{epstopdf}
\usepackage{ulem}

\newcommand{\beq}{\begin{equation}}
\newcommand{\eeq}{\end{equation}}
\newcommand{\bea}{\begin{eqnarray}}
\newcommand{\eea}{\end{eqnarray}}
\newcommand{\gae}{\lower 2pt \hbox{$\, \buildrel {\scriptstyle >}\over {\scriptstyle
\sim}\,$}} 
\newcommand{\lae}{\lower 2pt \hbox{$\, \buildrel {\scriptstyle <}\over {\scriptstyle
\sim}\,$}}

\begin{document}

\title[GRB radio afterglow rebrightening]{Radio rebrightening of the GRB afterglow by the accompanying supernova}

\author[Barniol Duran \& Giannios]{R. Barniol Duran$^{1}$\thanks{Email: rbarniol@purdue.edu, dgiannio@purdue.edu}, and D. Giannios$^{1}\footnotemark[1]$ \\
$^{1}$Department of Physics and Astronomy, Purdue University, 525 Northwestern Avenue, West Lafayette, IN 47907, USA}

\date{Accepted 2015 August 26; Received 2015 August 21; in original form 2015 April 11}

\pubyear{2015}

\maketitle

\begin{abstract}
The gamma-ray burst (GRB) jet powers the afterglow emission by shocking the surrounding medium, and radio afterglow can now be routinely observed to almost a year after the explosion.  Long-duration GRBs are accompanied by supernovae (SNe) that typically contain much more energy than the GRB jet.  Here we consider the fact that the SN blast wave will also produce its own afterglow (supernova remnant emission), which will peak at much later time (since it is non-relativistic), when the SN blast wave transitions from a coasting phase to a decelerating Sedov-Taylor phase.  We predict that this component will peak generally a few tens of years after the explosion and it will outshine the GRB powered afterglow well-before its peak emission. In the case of GRB 030329, where the external density is constrained by the $\sim 10$-year coverage of the radio GRB afterglow, the radio emission is predicted to start rising over the next decade and to continue to increase for the following decades up to a level of $\sim$ mJy. Detection of the SN-powered radio emission will greatly advance our knowledge of particle acceleration in $ \sim 0.1$c shocks.  
\end{abstract}


\section{Introduction}

The external shock, produced as the relativistic gamma-ray burst (GRB) jet interacts with the circumburst medium, is thought to give rise to the GRB afterglow radiation (e.g., \citealp{sarietal98, wijersandgalama99, panaitescuandkumar00}). Radio afterglows have been observed for up to a decade after the time of the explosion, when the GRB jet blast wave moves at subrelativistic speed. The total kinetic energy in the blast wave in GRBs is $\sim 10^{51}$ erg (after beaming correction is applied, see, e.g., \citealp{frailetal01}).  

Long-duration GRBs are associated with supernovae (SNe) of the rare broad-line Ic type (see, e.g, \citealp{woosleyandbloom06, hjorthandbloom12, melandrietal14}, and references therein).  These peculiar SNe have very large kinetic energies, which are $\sim 50$ times larger than the total (beaming-corrected) GRB energy, and their ejecta have also unusually high velocities (almost $\sim 0.1$c).  The SN ejecta will also interact with the external medium: it will also drive an external shock, which will produce an afterglow. This SN afterglow is simply the emission from the supernova remnant (SNR), therefore, we will refer to it as ``SNR emission." Since the total energy in the SN is much larger than the one in the GRB jet, the SNR emission will eventually outshine that of the GRB.  

In this work, we calculate the expected radio synchrotron emission from the SN external shock.  Due to the non-relativistic nature of the SN ejecta, it will only decelerate after a few tens of years.  As in the case of the GRB jet, the synchrotron emission from the SN blast wave depends on the external density and the shock microphysics, which are unknown quantities {\it a priori}; however, we will use our knowledge of GRB afterglow studies as a guideline. In particular, GRB 030329 has been followed up in the radio for a decade after the burst  and its light curve continues on a smooth, power-law decline implying constant particle acceleration efficiency when the shock becomes subrelativistic \citep{meslerandpihlstrom13}. We use the available sample of GRBs associated with SNe to predict the SNR radio emission.  We find that, within our assumptions, the SNR radio emission should be detectable, and lead to a rise of the radio light curves about a decade after the GRB trigger.

There are two long GRBs at low-enough redshift to allow for the optical detection of the accompanying SN at the same level of SNe accompanying GRBs (GRB-SNe).  However, for these two bursts no SN optical emission was detected \citep{dellavalleetal06, fynboetal06, galyametal06, gehrelsetal06}. These two cases imply that either no SN explosion took place or that the SN was optically faint, because, e.g., of a very low amount of $^{56}$Ni present in these explosions.  If the kinetic energy of the SNe of these two bursts was similar to that of other GRB-SNe, then we can use our model to predict their radio SNR emission.  This radio emission, if detected, would imply that a SN of the GRB-SNe type was present in these two GRBs.

This paper is organized as follows.  In Section \ref{Radio_afterglow}, we summarize the model we adopt for the synchrotron emission from the external shock of the GRB jet in the Sedov-Taylor regime.  In the same Section, we calculate the synchrotron emission expected from the external shock of the SN ejecta. In Section \ref{Application}, we take the known sample of GRBs with a SN association and calculate the expected SNR radio emission. We do the same with the two GRBs with no SN association mentioned above. We finish with a discussion in Section \ref{Discussion}.

\section{The radio afterglow at late times} \label{Radio_afterglow}

The GRB jet interacts with the external medium and drives a beamed blast wave that accumulates mass and decelerates.  As it reaches non-relativistic velocities, the blast wave enters the Sedov-von Neumann-Taylor (ST) phase (e.g., \citealp{taylor46}) and becomes approximately spherical (albeit with a displaced center from the center of the explosion, see Section \ref{SN_afterglow_section}).  Since we are interested in late time radio observations, we focus our discussion on the synchrotron emission in this phase. We present a short summary following the work of \citet{sironiandgiannios13} (see references therein, and, e.g., \citealp{chevalier82a, chevalier82b, chevalier98, liandchevalier99, chevalierandfransson06}, for a similar modeling of the dynamics and radio synchrotron emission of the SN blast wave as it interacts with the surrounding medium).

\subsection{The GRB jet afterglow}

For the GRB jet component, according to numerical studies (e.g., \citealp{wygodaetal11, decolleetal12, vaneertenandmacfadyen12}), the spherical ST solution can be used to estimate the observed flux after a time $t_{\rm ST}$, given by
\beq
t_{ST}  \approx 290 (E_{\rm GRB,51}/n_0)^{1/3} (1+z) \, \rm d.
\eeq
Beyond $t_{\rm ST}$, the blast wave radius, $R$, and velocity, $\beta$ (in units of the speed of light), are 
\bea
R &\approx& 9.4 \times 10^{17} (E_{\rm GRB,51}/n_0)^{1/5} t_{\rm yr}^{2/5} (1+z)^{-2/5} \, \rm cm ,
\label{R_ST} \\
\beta &\approx& 0.4 (E_{\rm GRB,51}/n_0)^{1/5} t_{\rm yr}^{-3/5} (1+z)^{3/5},
\label{beta_ST}
\eea
where $E_{\rm GRB}$ is the beaming-corrected kinetic energy of the GRB jet blast wave, $n$ is the number density of the external medium, $t$ is the observed time since the explosion, $z$ is the cosmological redshift, and we have used the common notation $Q_x = Q/10^x$ in c.g.s units. We assume here a constant density external medium, although a wind medium is also a possibility (see, e.g., \citealp{chevalierandli99}). At this stage, the synchrotron emission for an observed frequency $\nu$ (where max($\nu_a$,$\nu_m$) $< \nu < \nu_c$, and $\nu_a$, $\nu_m$ and $\nu_c$ are the synchrotron self-absorption, minimum injection and cooling frequencies, respectively) is 
\bea
F_{\nu} &\approx& (0.02 \, {\rm mJy}) {\bar{\epsilon}_{e,-1}}^{\, \, \,p-1} \epsilon_{B,-2}^{\frac{1+p}{4}} E_{\rm GRB,51}^{\frac{3+5p}{10}} n_0^{\frac{19-5p}{20}} \nonumber \\
&\times& t_{\rm yr}^{\frac{3(7-5p)}{10}} \nu_{\rm GHz}^{\frac{1-p}{2}} (1+z)^{\frac{5p-8}{5}} d_{\rm 27.5}^{-2},
\label{F_ST}
\eea
where $\bar{\epsilon}_e \equiv 4 \epsilon_e (p-2)/(p-1)$, $\epsilon_e$ ($\epsilon_B$) is the fraction of shocked fluid energy in electrons (magnetic field), $p$ is the power-law index of the energy distribution of the electrons and $d$ is the luminosity distance. The normalization corresponds to $p=2.4$, but it does not depend strongly on $p$. We assume $\epsilon_e$ and $\epsilon_B$ do not change with time.

As discussed in \citet{sironiandgiannios13}, caution must be taken when the minimum Lorentz factor of the electrons, $\gamma_{min}$, given by $\gamma_{min}  - 1 = (1/8) \bar{\epsilon}_e (m_p/m_e) \beta^2$, drops below $\sim 2$.  At this point, the system will transition to a new regime: the ``deep Newtonian" (DN) regime, where the spectrum of accelerated electrons can no longer be approximated as power-law of index $p$ in kinetic energy, but it still  follows a power-law distribution in momentum with slope $p$.  In this case, for $2<p<3$, the bulk of the electron energy is contributed by mildly relativistic particles with Lorentz factor of $\sim 2$ (see \citealp{granotetal06}). This happens when the velocity drops below 
\beq
\beta_{\rm DN} \approx 0.2 \, \bar{\epsilon}_{e,-1}^{\, \, -1/2},
\label{beta_DN}
\eeq
which occurs at a time $t_{\rm DN}$, see eq. (\ref{beta_ST}),
\beq 
t_{\rm DN} \approx 2.2 (E_{\rm GRB,51}/n_0)^{1/3} \bar{\epsilon}_{e,-1}^{\, \, \,5/6} (1+z) \, \rm yr,
\eeq 
and the synchrotron flux in this regime is given by 
\bea 
F_{\nu} &\approx& (0.01 \, {\rm mJy}) \bar{\epsilon}_{e,-1} \epsilon_{B,-2}^{\frac{1+p}{4}} E_{\rm GRB,51}^{\frac{11+p}{10}} n_0^{\frac{3+3p}{20}} \nonumber \\
&\times& t_{\rm yr}^{\frac{-3(1+p)}{10}} \nu_{\rm GHz}^{\frac{1-p}{2}} (1+z)^{\frac{4-p}{5}} d_{\rm 27.5}^{-2} .
\label{F_DN}
\eea

\subsection{The SNR emission} \label{SN_afterglow_section}

Long GRBs are (usually) accompanied by a powerful (and fast) SN, see, e.g., \cite{woosleyandbloom06, hjorthandbloom12}.  As the SN propagates through the external density, a SN blast wave will develop, which will also produce its own afterglow.  We refer to this signal the ``SNR emission".  In essence, it is similar to the GRB jet afterglow, but the SN blast wave is much slower (while the GRB jet is relativistic), and the SN energy (at least for SNe which accompany GRBs) is much larger than the typical (beaming-corrected) energy of GRB jets (see Table \ref{table1}).  

The geometry we have in mind is the following (see fig. 7 of \citealp{ramirezruizandmacfadyen10}).  Both the GRB blast wave (at late times) and the SN blast wave (always) are approximated as spherical blasts.  However, the GRB blast wave, because of its initial directionality, will be expanding as a (quasi-) sphere with a shifted center. This new center is located at a distance (from the central engine) at which the initially beamed GRB jet reaches a blast wave Lorentz factor of $\sim 2$. The SN blast wave, on the other hand, is a truly centered (on the central engine) explosion. It probes mainly the pristine ``unshocked" external density along the equatorial plane (not affected by the GRB jet blast wave) rather than the one in the poles.  Once the SN blast wave reaches a size similar to that of the GRB blast, it will outshine the latter.

\begin{table}
\begin{center}
\begin{tabular}{c|cccc}
\hline
GRB & SN & $z$ & $v$ [$10^3$ km/s] & $E_{\rm SN}$ [$10^{51}$ erg] \\
\hline
980425 &  1998bw & 0.0083 & 24 & $50 \pm 5$\\
030329 & 2003dh & 0.1685 & 29 & $40 \pm 10$\\
031203 & 2003lw & 0.106 & 21 & $60 \pm 10$\\
060218 & 2006aj & 0.0335 & 19 & $2 \pm 0.5$\\
100316D & 2010bh & 0.0593 &  & $\sim$10\\
101219B & 2010ma & 0.55 &  & \\
120422A & 2012bz & 0.283 & 20.5 & 41 \\
130702A & 2013dx & 0.145 &  24 & $35 \pm 10$\\
130427A & 2013cq & 0.34 & 32 & $64 \pm 7$\\
140606B & iPTF14bfu & 0.384 & & \\
\hline
\end{tabular}
\end{center}
\caption{GRB, its associated SN, redshift, the SN velocity at $\sim 10$ d after the explosion and the SN energy. GRB-associated SNe are typically very energetic, with several $\times 10^{52}$ erg, and fairly fast velocity of $\sim 0.1$c. In our modeling, for blank cells in this table we assume 1998bw-like values. Data from table 9.2 in \citet{hjorthandbloom12} (and references therein); additional data from \citet{sparreetal11,melandrietal12,xuetal13,levanetal14,schulzeetal14,deliaetal15,singeretal15}.} 
\label{table1}
\end{table}

\begin{figure}
\includegraphics[width=8.5cm, angle=0]{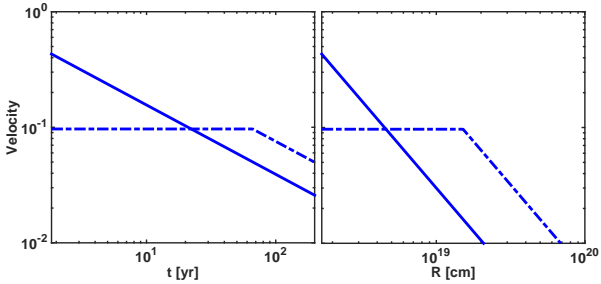}
\caption{Velocity (in units of the speed of light) as a function of observer time and blast wave radius. The solid (dot dashed) line marks the GRB jet (SN) component. The SN component has a constant velocity, until $t_{\rm dec,SN}$, when it decelerates. These lines correspond to the parameters of GRB 030329, described in the caption of Fig. \ref{fig2}. At $t \gae 20$ yrs, the SN blast overtakes that of the GRB jet and its emission dominates.}    
\label{fig1}
\end{figure}

The SN blast wave coasts with constant velocity until it starts to slow down after it has doubled its initial mass.  The deceleration time is  
\beq
t_{\rm dec,SN} \approx 29 \beta_{\rm SN,-1}^{-5/3} (E_{\rm SN,52.5}/n_0)^{1/3} (1+z) \, \rm yr,
\label{t_dec_SN}
\eeq
where $\beta_{\rm SN}$ and $E_{\rm SN}$ are the velocity and energy of the SN blast wave, respectively. At this time, the SN blast wave velocity will decrease as $\propto t^{-3/5}$, see eq. (\ref{beta_ST}).  The SN blast wave radius will increase linearly with time while the velocity is constant, but will increase slowly with time, as $\propto t^{2/5}$, after $t_{\rm dec,SN}$, see eq. (\ref{R_ST}).  Fig. \ref{fig1} shows the velocity of both components, the GRB and SN components, as a function of observer time and radius.

During the coasting phase, while the blast wave collects more and more external medium, the SNR emission light curve will rise rapidly as $\propto t^3$ (see below). Because of its subrelativistic speed $\beta_{\rm SN} \lae 0.1$, the SNR emission is in the DN regime, see eq. (\ref{beta_DN}).  For this reason, after $t_{\rm dec,SN}$, the SNR emission will decay as $\propto t^{-3(1+p)/10}$, as described in eq. (\ref{F_DN}), and the peak flux (at $t_{\rm dec,SN}$) will be given by
\begin{align}
F_{\nu} &(t=t_{\rm dec,SN}) \approx (44 \, {\rm \mu Jy}) \bar{\epsilon}_{e,-1} \epsilon_{B,-2}^{\frac{1+p}{4}} \beta_{\rm SN,-1}^{\frac{1+p}{2}} \nonumber \\
&\times E_{\rm SN,52.5} \, n_0^{\frac{1+p}{4}} \nu_{\rm GHz}^{\frac{1-p}{2}} (1+z)^{\frac{1-p}{2}} d_{\rm 27.5}^{-2}.
\label{F_DN_SN}
\end{align}
For $t<t_{\rm SN,dec}$, the SNR radio flux is given by
\begin{align}
F_{\nu} &(t<t_{\rm dec,SN}) \approx (2 \, {\rm \mu Jy}) \bar{\epsilon}_{e,-1} \epsilon_{B,-2}^{\frac{1+p}{4}} \beta_{\rm SN,-1}^{\frac{11+p}{2}} \nonumber \\
&\times n_0^{\frac{5+p}{4}} t_{\rm 10 yr}^{3} \nu_{\rm GHz}^{\frac{1-p}{2}} (1+z)^{\frac{-(5+p)}{2}} d_{\rm 27.5}^{-2}.
\end{align}
Old, nearby and fast moving SN ejecta, which interact with a dense external medium, would yield larger fluxes in this stage.

We can find a time, $t_{\rm eq}$, where the declining GRB afterglow and the rising SNR emission flux are equal. At these late times, the two blasts propagate at a similar speed and at $\sim$ pc scales away from the explosion center. Assuming constant density surrounding gas and the same microphysical parameters ($\epsilon_e$, $\epsilon_B$, and also $p$) we find
\bea
t_{\rm eq} &=& t_{\rm dec,SN} (E_{\rm GRB}/E_{\rm SN})^{1/3} \nonumber \\
&\approx& 9 \beta_{\rm SN,-1}^{-5/3} (E_{\rm GRB,51}/n_0)^{1/3} (1+z) \, \rm yr.
\label{t_eq}
\eea
In view of this last equation, determining observationally $t_{\rm eq}$ can actually provide an independent constrain on $E_{\rm GRB}$, $\beta_{\rm SN}$ and density (which does not depend on microphysics). This constrain depends most sensitively on the SN velocity.  On the other hand, if we have a good knowledge of $E_{\rm GRB}$ and $\beta_{\rm SN}$ we can determine the external density.  We note that for $t>t_{\rm SN,dec}$, the SNR emission exceeds that of the GRB by a factor $(E_{\rm SN}/E_{\rm GRB})^{(11+p)/10}\gg 1$, see eq. (\ref{F_DN}). For typical values, the flux ratio is $\gae100$.

The expressions presented in this Section (for both the GRB jet and SN components) are applicable when the radio observing frequency is max($\nu_a$,$\nu_m$) $< \nu < \nu_c$ (the cooling frequency is irrelevant in the radio band).  For the SN component, since the peak of the emission is produced when the blast wave is far away from the center of the explosion, and moves at a low velocity, it can be shown that both $\nu_a$ and $\nu_m$ are always below the observing radio frequency \citep{nakarandpiran11,sironiandgiannios13}. The same is usually true for the GRB jet component emission at very late times.

We note that the SN blast wave radius, $R_{\rm SN}$, at $t>t_{\rm dec,SN}$ increases as $t^{2/5}$, see eq. (\ref{R_ST}).  The angular size of the SNR emission at this stage would be $\theta_{\rm SN} \approx 2 R_{\rm SN}/d_A$, where $d_A=d/(1+z)^2$ is the angular diameter distance. The SN blast wave radius at this stage only depends on $E_{\rm SN}$, density and time, see eq. (\ref{R_ST}).  We can solve for density and substitute it in the SN blast wave synchrotron flux equation [eq. (\ref{F_DN}), but with $E_{\rm SN}$ instead of $E_{\rm GRB}$].  We find an interesting expression which does not explicitly depend on time since the explosion (although the angular size and flux do vary with time), as follows 
\bea
\bar{\epsilon}_{e,-1} \, \epsilon_{B,-2}^{\frac{1+p}{4}} &\approx& 30  \,
\bigg(\frac{F_{\nu}}{1 \, {\rm mJy}}\bigg)
\bigg(\frac{\theta}{1 \, {\rm mas}}\bigg)^{\frac{3+3p}{4}}  
 \, E_{\rm SN,52.5}^{-\frac{5+p}{4}} \nonumber \\
&\times& \nu_{\rm GHz}^{\frac{p-1}{2}} (1+z)^{-(2+p)} d_{\rm 27.5}^{\frac{11+3p}{4}} .
\label{epsiloneepsilonB}
\eea
Strictly speaking the normalization is only valid for $p \approx 2.4$.  Therefore, by measuring at a given time the flux and size after the peak of the SN radio light curve, we can constrain $\sim \epsilon_e \epsilon_B$ with our knowledge of $E_{\rm SN}$. 

\section{Application to GRBs} \label{Application}

\subsection{Sample of GRBs}

We present the current sample of GRBs for which there is a strong association with SNe, see Table \ref{table1}.  We provide the approximate velocity of the SN (at $\sim 10$ d) and the estimated total energy $E_{\rm SN}$. With this information, we can predict the SNR light curves for a given external density and microphysical parameters. 

Alternatively, we can model the very late radio GRB afterglow data ($\sim$ years time-scale) in the context of synchrotron emission from the GRB jet blast wave and determine the density and microphysical parameters.  The medium that is probed by the blast wave at such long time-scales (large distances), and the microphysics of the blast wave when the velocity is small, would be a good proxy of the medium/microphysics also expected for the SNR emission.  This exercise is only possible in the case of GRB 030329 for which there is radio afterglow data for $\sim 8$ yrs after the burst \citep{meslerandpihlstrom13}.  We will treat this case separately.  In general, GRB afterglows are detectable for less than a year. In these cases, the external density probed might be very different than the one probed by the SN blast wave at very large distances (the same applies to the microphysics), and any attempts made to connect the early GRB radio emission with the late SNR one would be plagued by this uncertainty.  For this reason, for these GRBs we will leave the density/microphysics as free parameters when we predict the SNR emission.  We will also treat GRB 980425 separately, since it is the closest GRB observed to date. We will then consider the rest of GRBs in Table \ref{table1} for our analysis.

In addition to the GRBs in Table \ref{table1}, we will also consider two GRBs for which a SN was expected but not found, GRB 060505 and GRB 060614, at $z=0.0894$ and $z=0.125$, respectively \citep{dellavalleetal06, fynboetal06, galyametal06, gehrelsetal06}.  For these two, we will use 1998bw as a template to predict the expected SNR emission, which could serve as a test whether or not a SN like 1998bw was truly present in these two explosions.

We note that throughout this paper we set an approximate limit of detection of a radio signal of $\sim 20 \mu$Jy, which is expected by EVLA at 4.9 GHz \citep{perleyetal11}.  We will use this limit to determine if and when the SNR emission will be observable.  

\subsection{GRB 030329} \label{030329_section}

\begin{figure}
\includegraphics[width=8.5cm, angle=0]{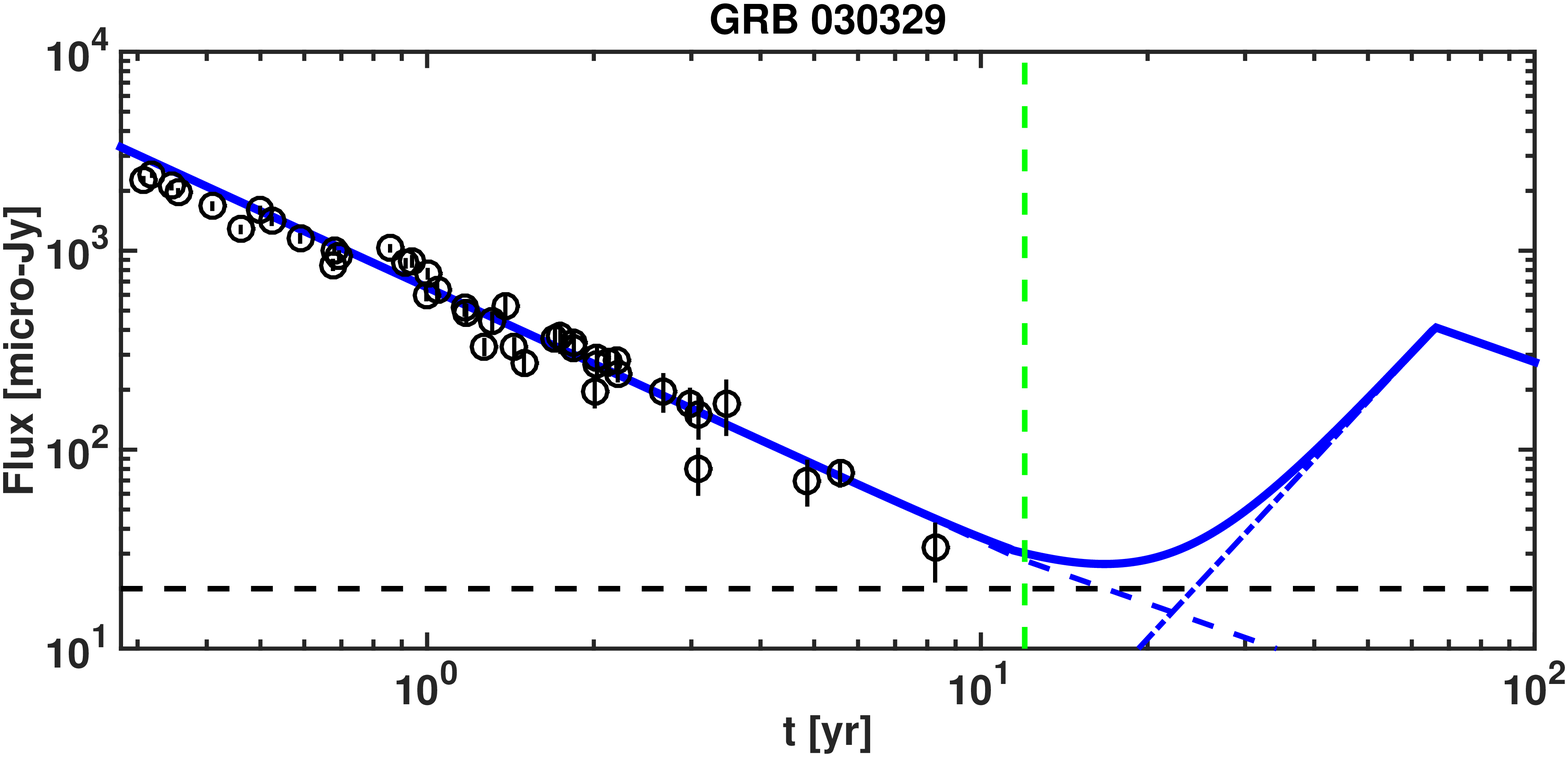}
\includegraphics[width=8.5cm, angle=0]{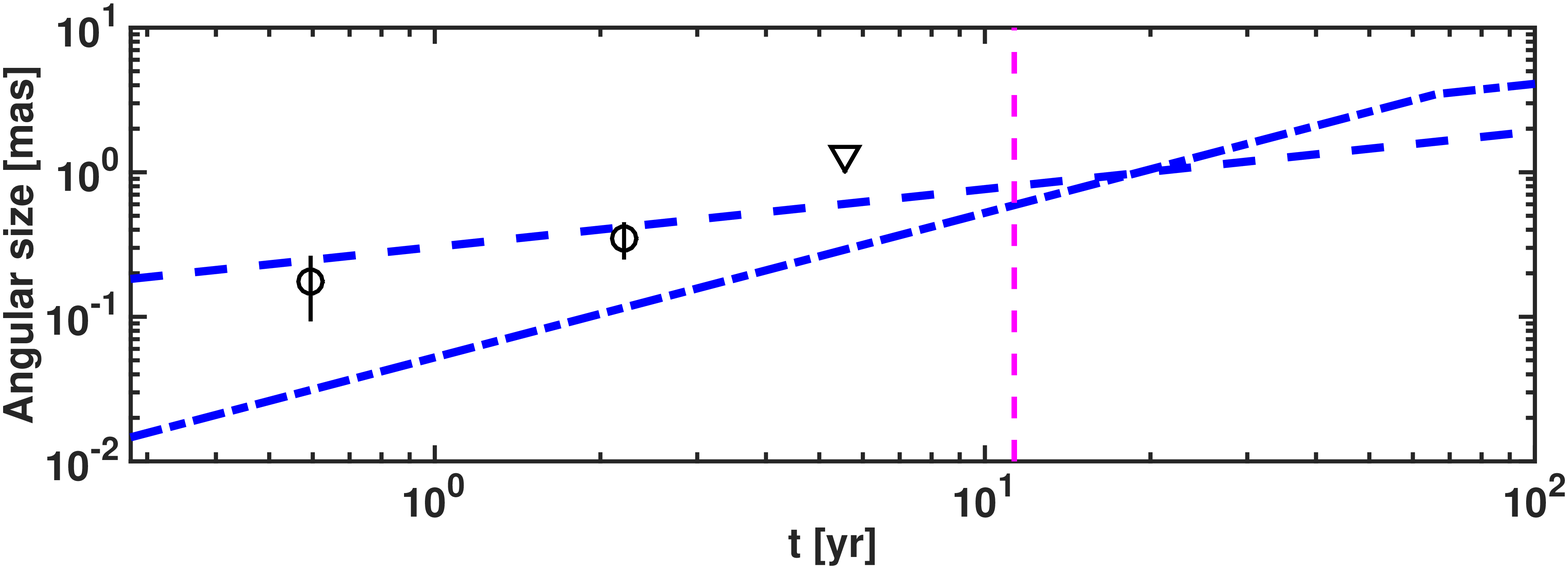}
\caption{{\it Top panel:} We fit the radio 4.9 GHz light curve (black circles, \citealp{mesleretal12, meslerandpihlstrom13}, and references therein) with the {\it lowest} possible density allowed by the data of $n = 0.2$ cm$^{-3}$ (dashed blue line), see Section \ref{030329_section}. With these parameters, we predict that the SNR radio emission (dot dashed blue line) should rise above the EVLA flux limit of $\sim 20 \mu$Jy (horizontal black dashed line) by 2030 {\it at the latest} (blue solid thick line marks the sum of both the GRB jet and SNR emission components).  The year 2015 is marked with a vertical green dashed line. Strictly, speaking, our fit to the radio data with eq. (\ref{F_ST}) is valid only for $t>t_{\rm ST} \sim 2$ yrs (same applies to the bottom panel). We only include radio data with $t>100$ d for both panels. {\it Bottom panel:}  Measured angular sizes (black circles) and upper limit (triangle) of the GRB 030329 radio afterglow source (\citealp{mesleretal12} and references therein).  The blue dashed (dot dashed) thin line is the GRB (SN) afterglow angular size for the model presented in the top panel. The expected angular size at a given time is the maximum between both components.  The dashed magenta vertical line marks when the ``Deep Newtonian" regime starts for the GRB component (the SN component is always in this regime).}    
\label{fig2}
\end{figure}

GRB 030329 holds the record for the longest radio afterglow ever detected.  The last reported data point is at $\sim 8$ yr \citep{meslerandpihlstrom13}.  We can fit the late radio data, which decays as $t^{-1.27}$,  using the synchrotron flux of the GRB jet blast wave, see eq. (\ref{F_ST}) (which points out to a power-law index of electrons $p\approx 2.2$). We can use the same density/microphysical parameters used to model the radio GRB afterglow to predict the SNR emission, since at such long observed time-scales both the GRB and the SN components move at similar velocities  and probe the density of the external medium at similar scales.  Several groups have fitted the radio afterglow (see, e.g., \citealp{mesleretal12, meslerandpihlstrom13}, and references therein), and a few sets of density/microphysical parameters have been provided.  Fits point out to a uniform external medium of $\sim 1$ cm$^{-3}$.  

As can be seen from eqs. (\ref{t_dec_SN}) and (\ref{F_DN_SN}), the larger the density, the earlier and brighter the SNR emission peak will be. For this GRB we take a conservative approach and estimate the {\it lowest} density allowed by the data.  We do this by setting the maximum values of the microphysical parameters ($\epsilon_e=\epsilon_B = 1/3)$ and we set the energy to the largest value found for this GRB of $1.5\times10^{51}$ erg, see tables 2 and 3 of \citet{meslerandpihlstrom13}.  For a fixed observed flux, these choices will yield the minimum required density, see eq. (\ref{F_ST}), which is $n \approx 0.2$ cm$^{-3}$.  This constrains the rise of the SNR radio emission to occur on or before $\sim$2030, see Fig. \ref{fig2}. We also include the afterglow size (both the GRB and SN components) for the same model.  As can be seen, the SNR emission size dominates the GRB afterglow size $\sim 20$ yr after the explosion.  

Using the last observed radio point at $\sim 8$ yr as a limit of when the rise of the SNR emission occurred, e.g., $t_{\rm eq} \gae 8$ yr, we can find an upper limit on the external density (for a given $\beta_{\rm SN}$ and $E_{\rm GRB}$), see eq. (\ref{t_eq}).  For the parameters mentioned above, we find that the density cannot be larger than $\sim 4$ cm$^{-3}$, otherwise we would have seen the rise of the SNR emission already.

\subsection{GRB 980425}

\begin{figure}
\includegraphics[width=8.5cm, angle=0]{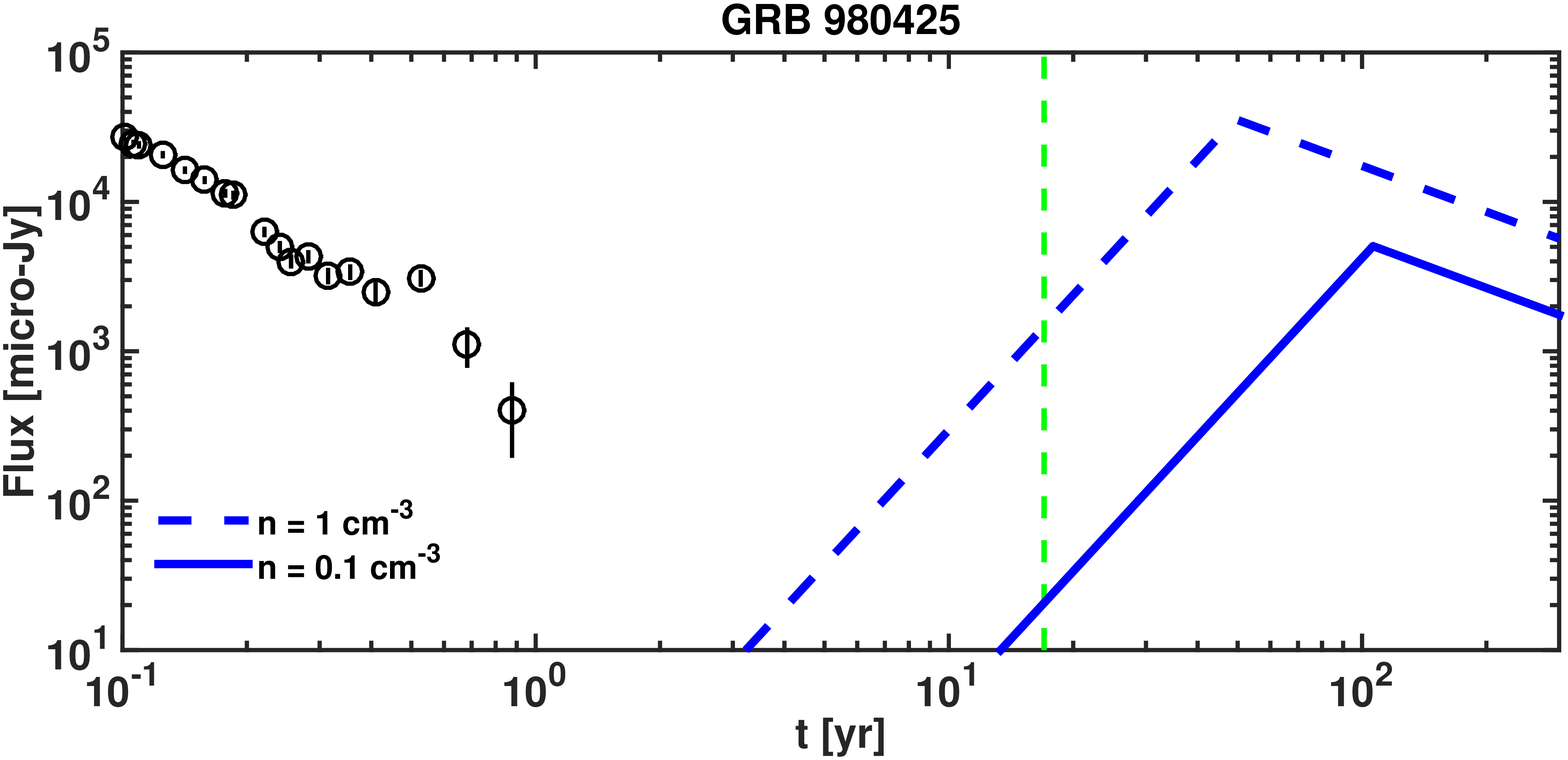}
\includegraphics[width=8.5cm, angle=0]{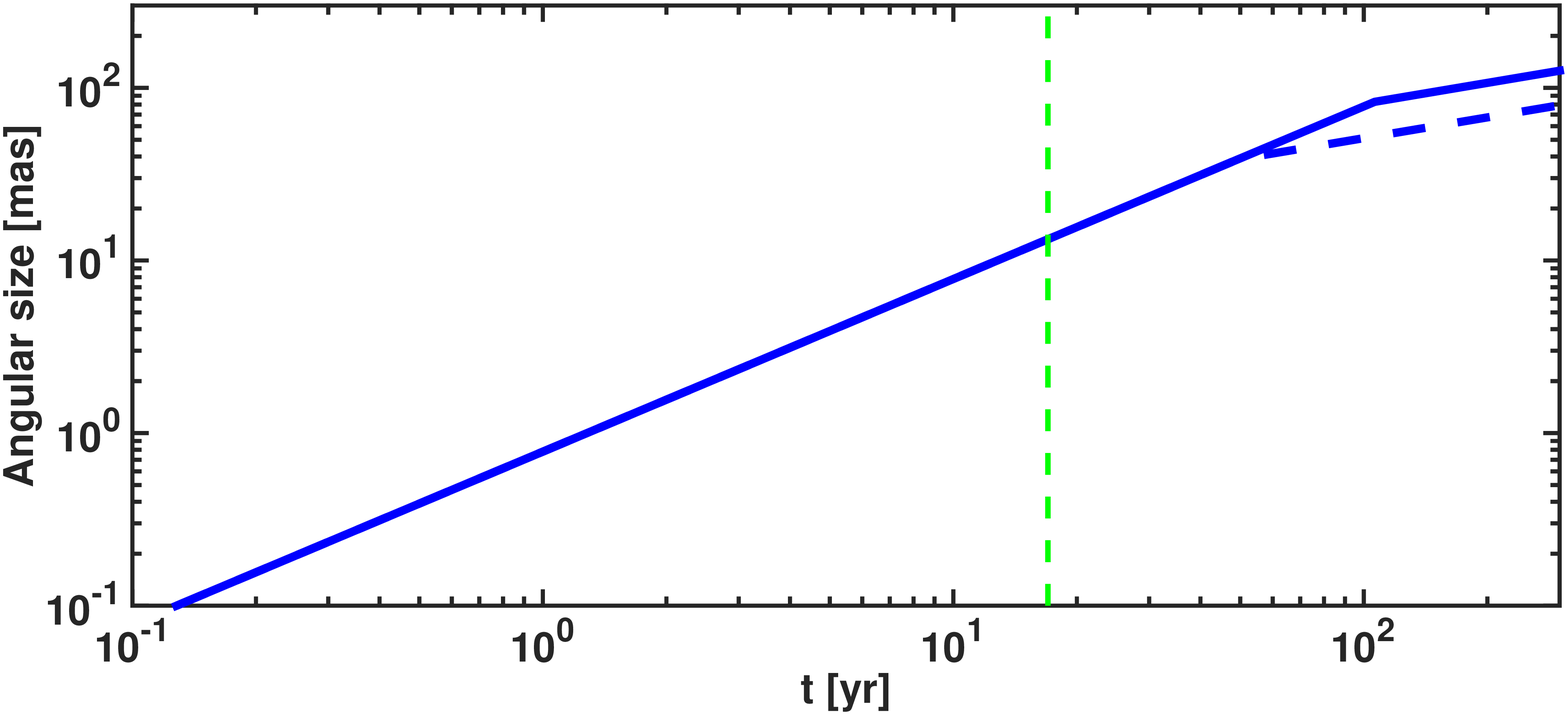}
\caption{{it Top panel:} The 4.8 GHz radio observations of GRB 980425 are denoted with black circles \citep{kulkarnietal98, frailetal03}. We use $\epsilon_e=0.2$, $\epsilon_B=0.01$ and $p=2.4$ to calculate the SNR emission, with $n=1$ cm$^{-3}$ and  $n=0.1$ cm$^{-3}$ (dashed and solid blue lines, respectively). The vertical green dashed line marks the year 2015. {\it Bottom panel:} The size of the SN blast wave for the same densities presented above; solid and dashed lines overlap before $\sim 50$ yrs, since the coasting velocity is kept fixed, and only the density is changed.}    
\label{fig3}
\end{figure}

We now focus on GRB 980425.  For this GRB, there is no radio data after about a year, therefore, it is not possible to follow the same procedure as for GRB 030329.  To calculate the SNR radio emission for this burst we simply assume a typical value found for GRB afterglows $\epsilon_e \approx 0.2$, and an optimistic value of $\epsilon_B \approx 0.01$ (\citealp{barniolduran14, santanaetal14}, see also simulations by \citealp{sironiandspitkovsky11}).  We also use $p=2.4$.  We use these values of $\epsilon_e$, $\epsilon_B$ and $p$ throughout the rest of the paper, unless otherwise noted. We allow for two possible values for the external density, $n=0.1$ cm$^{-3}$ and a more optimistic value of $n=1$ cm$^{-3}$. Using the velocity and energy of the SN component, we can predict the SNR emission, and it is shown in Fig. \ref{fig3}.  As mentioned above, a higher (lower) density yields a SNR emission that peaks earlier (later) and is brighter (weaker), but the shape of the afterglow is preserved.  For the parameters adopted for this SNR emission, it will reach values at the same level of the GRB radio afterglow detected $< 1$ yr after the explosion.

We use GRB 980425 as a benchmark case.  Assuming an optimistic density of 1 cm$^{-3}$, we find that a GRB with a 1998bw-like SN should be located at a distance $\lae 1$ Gpc ($z \lae 0.25$), so that its SNR radio emission peaks above $\sim 20 \mu$Jy. 

\subsection{Rest of the sample}

\begin{figure}
\includegraphics[width=8.5cm, angle=0]{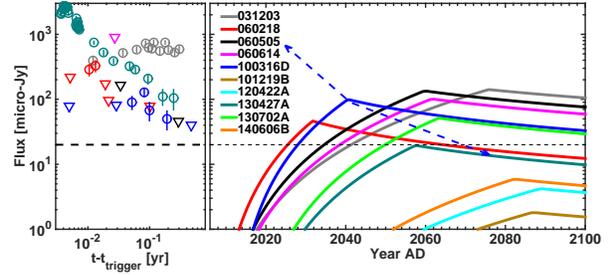}
\caption{The left-hand panel shows the observed radio data (circles) and upper limits (triangles) as a function of time since the trigger (in log-scale) for some bursts in our sample (see legend for colors). The right-hand panel shows the predicted SNR emission as a function of absolute observing time in years (in linear scale). For these light curves we use $\epsilon_e=0.2$, $\epsilon_B=0.01$, $p=2.4$, and an external constant density of $n=1$ cm$^{-3}$.  The velocity and energy of the SN component can be found in Table \ref{table1} and are used to obtain the SNR emission light curves. For a higher (lower) density of $n=10$ cm$^{-3}$ ($n=0.1$ cm$^{-3}$), the SNR emission will peak at an earlier and higher (later and weaker) flux, indicated by the blue dashed arrow (see GRB 100316D as an example). The horizontal black dashed line marks the EVLA flux limit of $\sim 20 \mu$Jy. For GRB 060505, and 060614, we assume a 1998bw-like SN blast wave. Radio data (presented only for some bursts) and predictions at: 4.9 GHz for GRB 032103 \citep{soderbergetal04}, GRB 060218 \citep{soderbergetal06a}, GRB 060614 \citep{londishetal06}, GRB 130427A \citep{vanderhorstetal14}, GRB 130702A \citep{vanderhorst13}; 5.4 GHz for GRB 100316D \citep{marguttietal13}; 6.1 GHz for GRB 140606B \citep{singeretal15}; 8.5 GHz for GRB 060505 \citep{ofeketal07}, GRB 101219B \citep{frail11}, and 15 GHz for GRB 120422A \citep{schulzeetal14}.}
\label{fig4}
\end{figure}

For the rest of the GRBs in Table \ref{table1}, we follow a similar procedure as the one followed for GRB980425.  For GRB 101219B and 140606B, we assume a 1998bw-like SN. For GRB 100316D, we assume a 1998w-like SN velocity. We predict their SNR radio emission and show them in Fig. \ref{fig4}, along with their available late time radio data.  We also include the two GRBs with no SN identification (GRB 060505 and 060614) assuming that a SN blast wave with the properties of 1998bw was present. For the parameters adopted in Fig. \ref{fig4}, it can be seen that for some bursts the SNR radio emission could potentially be detected, while for others it is too weak to be detected with current instruments.

As the external density is an uncertain parameter, we allow for a larger (smaller) density than the adopted value of $1$ cm$^{-3}$ by a factor of $10$.  We indicate the position of the peak SNR emission for these different densities with a dashed arrow in Fig. \ref{fig4}.  We do this only for the case of GRB 100316D, which gives a sense of the density dependence of our calculation.

\subsection{Host galaxy radio contribution}

The host galaxy contribution to the observing radio band is a critical factor to be able to identify the SNR radio emission a few tens of years after the burst.  For example, the host galaxy contribution of GRB 980425 appears to be bright, on the order of $420 \pm 50$ $\mu$Jy at 4.8 GHz \citep{michalowskietal09}.  Also, the host galaxy contribution of GRB 031203 is estimated to be $216 \pm 50$ $\mu$Jy at 5.5 GHz (\citealp{stanwayetal10}, see, also, \citealp{soderbergetal04}).  Other bursts in our sample have a host contribution in radio which is smaller: 3-$\sigma$ limits at 5.5 GHz of $117$, $37$ and $33$ $\mu$Jy for GRB 060218, GRB 060505, GRB 0606014, respectively \citep{stanwayetal10,michalowskietal12}. Therefore, setting a precise baseline observation to reduce the errors in the host contribution measurements is of utmost importance to be able to detect the radio SNR emission.

\section{Discussion} \label{Discussion}

As the GRB jet interacts with the external medium, it produces an external shock, which gives rise to the observed GRB afterglow emission.  The SN ejecta, which accompanies long GRBs, will also interact with the external medium and will also drive an external shock.  We have calculated the expected synchrotron emission from this shock and predicted the early phases of the radio SNR emission for the current sample of GRB associated with SNe.  

The prospects for detection depend on the assumed values of the external medium and microphysics of the SN blast wave, which are a priori unknown and are the free parameters in our model.  At least for the case of GRB 030329, where the GRB radio afterglow was detected to about $\sim 10$ yrs after the explosion, we have estimates for the external density at large distances from the explosion (parsec scale), and also on the microphysics of a blast wave that moves close to $\sim 0.1$c, and we can use these to calculate the SNR emission. For this burst, the prospects for detection are optimistic. However, if the SN ejecta shock microphysical values are much different than those for the GRB shock, then the expected SNR radio emission would be much harder to predict, since the peak of the SNR emission light curve depends linearly (almost linearly) on the fraction of shocked energy that goes into electrons (magnetic field).  

The value of the power-law index of electron energy distribution expected in non-relativistic shocks is $p \approx 2$ (e.g., \citealp{blandfordandeichler87}); however, we use a slightly larger conservative value of $p \approx 2.2-2.4$.  We also assume that $\epsilon_e$ is of the order of 10 per cent, guided by GRB afterglow studies and by the GRB afterglow fits done for GRB 030329.  It is possible that $\epsilon_e$ is smaller for a non-relativistic blast wave, which will weaken the SNR radio signal significantly (linearly, as mentioned before).  Since the magnitude of $\epsilon_e$ for subrelativistic shocks continues to be actively studied (e.g., \citealp{parketal14}), we take the value adopted in this paper as an optimistic one.  Nevertheless, SNe are well-known active sites of particle acceleration and radio emission. In fact, mildly relativistic SNe, which radiate in the radio band, have been detected even without the presence of a GRB; ``typical" Ibc SNe efficiently accelerate particles in $\sim 0.1$c shocks, which produce radio emission (e.g., \citealp{soderbergetal10}).  Thus, expecting a radio signal from SNe accompanying GRBs is a natural extension of what we presently know, and would provide an excellent opportunity to study the behavior of $\epsilon_e$ as the shock transitions to $\sim 0.1$c. 

The radio SN rebrightening and the large size of the SN blast wave provide a unique opportunity to directly resolve these sources, if the GRB is located at $z \lae 0.2$.  In the coming years, GRB 030329 and 980425 may reach flux levels in excess of 100 $\mu$Jy and source size of $> 1$ mas ($\sim 20$ mas for GRB 980425 / SN 1998bw), well-within the resolving capabilities of current instruments.

We have also assumed that the external density probed by the SN blast wave, at $\sim$ several pc from the center of the explosion, is constant and roughly $\sim 1$ cm$^{-3}$ (although we have considered also values larger and smaller than this by a factor of 10).  To explore these assumptions, let us assume a GRB progenitor Wolf-Rayet star with wind velocity of $\sim$ 1000 km/s and mass loss rate of $\sim 10^{-5} M_{\odot}$ yr$^{-1}$ that extends to infinity.  At parsec scales, the wind density drops to $\sim 0.03$ cm$^{-3}$.  It is safe to assume that at some point, this density will encounter a ``floor."  If the star is in isolation and moving at a typical velocity of $v_{\rm star} \sim 100$ km/s with respect to the external medium, the floor of interstellar density value of $\sim 1$ cm$^{-3}$ will be limited by the ram pressure of the wind against the external medium, and will be set at a distance of \citep{fryeretal06}
\bea
r_{\rm wind} &=& 1.8 {\rm pc} \, 
\bigg(\frac{\dot{M}_{wind}}{10^{-5} M_{\odot} {\rm yr}^{-1}}\bigg)^{0.5}
\bigg(\frac{v_{wind}}{1000 \, {\rm km \, s^{-1}}}\bigg)^{0.5} \nonumber \\
&& \times \bigg(\frac{n}{1 \, {\rm cm}^{-3}}\bigg)^{-0.5}
\bigg(\frac{v_{\rm star}}{100 \, {\rm km \, s^{-1}}}\bigg)^{-1}.
\eea
Since the SN blast wave decelerates at large distances $\gae r_{\rm wind}$, it is safe to assume that it already probes the ISM medium, and not the wind from its progenitor. 

If the star is not in isolation, but part of a young stellar cluster, then taking the Galactic center clusters as an example, which contain 10-15 per cent of all Wolf-Rayet stars in the Galaxy, we can find a typical density as the winds of several stars interact with each other (see, e.g., \citealp{mimicaandgiannios11}).  Let us take a cluster of $R_c \sim 1$ pc size with $N\sim100$ O-stars, with typical winds of $\dot{M}_{\rm wind} \sim 10^{-6} M_{\odot}$ yr$^{-1}$ and $v_{\rm wind} \sim$ 1,000 km/s \citep{figer04}.  The stellar separation is $d \sim R_c N^{-1/3}$.  At the place were winds of different stars interact with each other, the density is typically of order $1$ cm$^{-3}$. This can be obtained by calculating the typical wind density at a distance $\sim d$, which is $\dot{M}_{\rm wind}/(4 \pi d^2 v_{\rm wind})$ or, alternatively, by adding the winds of all stars in the cluster, which yields a total wind density of $N \dot{M}_{\rm wind}/(4 \pi R_c^2 v_{\rm wind})$, both expressions yield $\sim 1$ cm$^{-3}$ density.  Although the density as winds interact will not be constant, $n \sim 1$ cm$^{-3}$ is a fair guess for the characteristic density of the ambient gas\footnote{The ISM may be clumpy (as revealed in, e.g., Galactic remnants), and that makes the blast appearance much more rich (see, e.g., 
\citealp{obergaulingeretal14}). Since we are interested in the average properties of a structure that is hardly resolved, a constant density medium may be a good first approximation.}.  In the unlikely case that this constant density floor is not reached and the wind-like medium extends to hundreds of pc scales, then the SN blast wave will decelerate $\gae 5000$ yr after the explosion for the typical parameters considered above. In this case, the strong and distinct radio SNR rebrightening discussed in this paper will not occur.

In this paper we have taken a simplistic approach and completely separated both components: the GRB and the SN ejecta.  It is possible that both components are ``connected" in the energy-velocity space, and there is a continuum of components between the relativistic GRB one and the non-relativistic SN one, which could be tested with future observations. However, this seems not to be the case for GRB 030329. For this burst, it appears that a single component is enough to model the GRB radio afterglow, and the presence of another component could have already potentially been observed.  If a continuum of components exists between the GRB and the SN one, then energy would continuously be injected to the blast wave while it decelerates.  This may also produce a distinct rebrightening in the light curve.  For a constant density medium, as the energy injection transitions to the non-relativistic phase, the decaying radio light curve is expected to start rising as long as the energy is injected steeper than $\propto \beta^{-2.5}$, where $\beta$ is the blast wave velocity\footnote{The energy-velocity of the GRB and SN components in ``regular" GRBs appears to be flatter than this condition (e.g., \citealp{marguttietal13}); however, already GRB 030329 points out that the two components might not be connected, but otherwise be well-separated.} [this statement is somewhat dependent on the precise radio spectrum, see eq. (23) in \citealp{barniolduranetal15}].  In addition, the rise in the light curve would also appear if the energy injection is accompanied by a transition of the external medium from a wind-like to a constant density one.

Nonetheless, a connection in energy-velocity space between the SN and GRB components is expected for low-luminosity GRBs ({\it ll}GRB) in our sample, where the GRB emission is likely produced in the shock breakout scenario (e.g. \citealp{kulkarnietal98, matznerandmckee99, tanetal01, campanaetal06}, and recently, \citealp{nakarandsari12}). Here, the explosion energy is deposited at the center of the progenitor and this drives a shock that crosses the star and accelerates at the stellar edge.  This acceleration dictates a specific relation between the fast and slow moving material \citep{matznerandmckee99}, which is seen in regular SNe, but not in {\it ll}GRB, where there seems to be more energy in the fast moving ejecta (e.g., \citealp{soderbergetal06a, marguttietal13, barniolduranetal15}). Recently, \cite{barniolduranetal15} modeled the $\lae 1$ yr radio afterglow data of {\it ll}GRBs within the relativistic shock breakout model.  While it is tempting to use their model (and their results) to predict the SNR radio emission, there are a few reasons why we have decided not to use it.  First, as mentioned before, the $<1$ yr afterglow data probes the external density at distances much closer to the source than the larger distances probed by the SN component.  Secondly, since at face value, an extrapolation in energy-velocity space for these sources does not work, extrapolating this model all the way to $\beta_{\rm SN}$ and $E_{\rm SN}$ would be dangerous.  Finally, \cite{nakar15} has recently suggested that the SN component is deposited at the center of the explosion, whereas the fast moving material is deposited by a (failed) GRB jet, thus decoupling both components.  For these reasons, even for the {\it ll}GRBs in our sample, we use the conservative approach to clearly separate both components.  Nevertheless, even in models where the interaction of the SN ejecta with the surrounding medium is invoked to explain the $<1$ yr GRB radio data (e.g., \citealp{liandchevalier99}, for GRB 980425), a radio rebrightening at late times will still occur.  Here, the SN blast wave velocity is initially found to be $\gae 0.5$c, and the energy $\sim 10^{50}$ erg, whereas $>100$ times more energy is carried by the $\sim 0.1$c--moving material, which will decelerate at much later times and can power a distinct peak in the emission surpassing (long after the explosion) the emission of the GRB afterglow or other mildly relativistic components (unless the wind-like medium found for GRB 980425 extends to very large distances -- see discussion above).     

For simplicity, we have not included the contribution from the GRB ``counter jet" in our calculations: the second GRB jet that points away from us. Although there could be some confusion between the possible re-brightening of the counter jet and the SNR radio emission, the energy in the SN component is much larger than that of the counter jet, making the SNR emission much brighter.  Also, the SNR radio emission is expected to peak at a few tens of years after the explosion, whereas the counter jet contribution to the GRB radio afterglow should peak in a few years time-scale (\citealp{granotandloeb03}, see also \citealp{sironiandgiannios13}, and simulations by \citealp{vaneertenetal10}). Even in the case of a wind medium, the GRB counter jet should peak before $\sim 10$ yrs \citep{decolleetal12}. For the case of GRB 030329, there is no clear evidence yet for a re-brightening due to the counter jet (e.g., \citealp{vanderhorstetal08, mesleretal12}).

There are several regular type Ibc SNe (not associated with GRBs) with upper limits on their radio emission at late times after the explosion \citep{soderbergetal06b}.  These SNe usually have kinetic energy of $\sim 10^{51}$ erg with velocity of 10,000 km/s. As a result of the low expansion velocity (of this particular component -- see below), these SNe are expected to be fainter radio emitters. For these SNe, we can also predict the expected SNR radio emission.  In the sample of \cite{soderbergetal06b}, SN1985F provides the strongest constraint since it is a nearby source (7.7 Mpc) with a strong upper limit ($37 \mu$Jy at 8.46 GHz, $\sim 18$ yrs after the explosion). This constrains the external density for this source to be $\lae 0.7$ cm$^{-3}$.  Weaker density constraints can be determined for other SNe in this sample.  On the other hand, typical Type Ibc SNe do show bright radio emission that peaks on $\sim$ tens of days time-scale; however, this emission is produced by a $\sim 0.1$c--component with kinetic energy of $10^{46}-10^{48}$ erg.

In addition to predicting the SNR radio emission for GRBs with SNe association, we have also predicted it for GRBs with no optical SNe identification to very strong limits. The detection of a SNR radio emission would settle if a 1998bw-like event was truly present in these explosions. Similar predictions could be done for short GRBs at low redshifts, for which an accompanying SN is not expected and is not observed either to strong limits.  For example, if a 1998bw-like event accompanied the short GRB 080905A, which is at a redshift similar to GRB 060614, then a similar SNR radio emission than the one calculated for GRB 060614 (see Fig. \ref{fig4}, just shifted in time by $\sim 2$ yr) would be present. Similar predictions could be made for other nearby short GRBs (see fig. 2 in \citealp{berger14}).  

The detection of the SNR radio emission would help to constrain the particle acceleration  and magnetic field generation mechanisms for $\lae 0.1$c shocks.  It will also allow us to constrain the external medium few pc scale from the center of the GRB explosion. Even if it turns out that the SNR radio emission is not observed, it will inform us that, for some reason, particles are not accelerated in $\lae 0.1$c shocks, and this would help guide current studies of particle acceleration models. We strongly encourage regular radio follow-ups of the locations of GRB 980425 and 030329. Any burst with $z<0.2$ that is more than one decade old also makes a promising target for radio follow-up. 

\section*{Acknowledgements}

We thank Paz Beniamini, Petar Mimica, Enrico Ramirez-Ruiz, Lorenzo Sironi and Alexander van der Horst for useful comments on the manuscript. We acknowledge support from NASA grant no. NNX13AP13G.



\end{document}